\documentclass[twoside]{article}
\usepackage{qic,epsfig}

\usepackage{tikz}
\usepackage{amsmath}
\usepackage{amssymb}  
\usepackage{amsthm}
\usepackage{amsfonts}
\usepackage{graphicx}
\usepackage{bbm}
\usepackage{url}

\usepackage{braket}
\usepackage{ upgreek }
\usepackage{dsfont}

\usepackage{authblk}
\usepackage{algorithm}
\usepackage{algpseudocode}
\usepackage{pifont}
\usepackage{afterpage}
\usepackage[plain]{fancyref}
\usepackage[title]{appendix}

\usetikzlibrary{decorations.pathmorphing}
\usetikzlibrary{positioning}
\usetikzlibrary{decorations.text}

\usepackage{hyperref}
\hypersetup{
    colorlinks,
    citecolor=black,
    filecolor=black,
    linkcolor=black,
    urlcolor=black
}

\theoremstyle{plain}               
\newtheorem{thm}{Theorem}[section]
\newtheorem{lem}{Lemma}[section]
\newtheorem{cor}{Corollary}[section]

\newtheorem{defn}{Definition}[section]

\theoremstyle{remark}

\def\beq{\begin{equation}}
\def\eeq{\end{equation}}
\def\bq{\begin{quote}}
\def\eq{\end{quote}}
\def\ben{\begin{enumerate}}
\def\een{\end{enumerate}}
\def\bit{\begin{itemize}}
\def\eit{\end{itemize}}

\def\lb{\left(}
\def\rb{\right)}

\def\l|{\left|}
\def\r|{\right|}

\newcommand\C{\mathbbm{C}}

\newcommand\R{\mathbbm{R}}
\newcommand\N{\mathbbm{N}}
\newcommand\M{\mathcal{M}}
\newcommand\D{\mathcal{D}}

\newcommand{\ketbra}[1]{|#1\rangle\langle#1|}
\newcommand{\tr}{\text{tr}}
\newcommand{\one}{\mathds{1}}
\newcommand{\id}{\text{id}}

\newcommand{\gibbs}{\frac{e^{-\beta H}}{\mathcal{Z}_\beta}}
\newcommand{\bigO}[1]{\mathcal{O}\lb#1\rb}

\newcommand*{\fancyrefthmlabelprefix}{thm}
\newcommand*{\fancyreflemlabelprefix}{lem}
\newcommand*{\fancyrefcorlabelprefix}{cor}
\newcommand*{\fancyrefdefilabelprefix}{defi}
\frefformat{plain}{\fancyreflemlabelprefix}{lemma\fancyrefdefaultspacing#1}
\Frefformat{plain}{\fancyreflemlabelprefix}{Lemma\fancyrefdefaultspacing#1}
\frefformat{plain}{\fancyrefthmlabelprefix}{theorem\fancyrefdefaultspacing#1}
\Frefformat{plain}{\fancyrefthmlabelprefix}{Theorem\fancyrefdefaultspacing#1}
\frefformat{plain}{\fancyrefcorlabelprefix}{corollary\fancyrefdefaultspacing#1}
\Frefformat{plain}{\fancyrefcorlabelprefix}{corollary\fancyrefdefaultspacing#1}
\frefformat{plain}{\fancyrefdefilabelprefix}{definition\fancyrefdefaultspacing#1}
\Frefformat{plain}{\fancyrefdefilabelprefix}{Definition\fancyrefdefaultspacing#1}
\Frefformat{plain}{\fancyrefdefilabelprefix}{algorithm\fancyrefdefaultspacing#1}

\begin{document}
\setlength{\textheight}{8.0truein}    

\runninghead{Perfect Sampling for Quantum Gibbs States}
            {Daniel Stilck Fran\c{c}a}

\normalsize\textlineskip
\thispagestyle{empty}
\setcounter{page}{1}


\vspace*{0.88truein}

\alphfootnote

\fpage{1}

\centerline{\bf
PERFECT SAMPLING FOR QUANTUM GIBBS STATES}

\vspace*{0.37truein}
\centerline{\footnotesize
Daniel Stilck Fran\c{c}a}
\vspace*{0.015truein}
\centerline{\footnotesize\it Department of Mathematics, Technische Universit\"at M\"unchen, Boltzmannstrasse 3}
\baselineskip=10pt
\centerline{\footnotesize\it Garching, 85748, Germany}
\vspace*{10pt}

\vspace*{0.21truein}

\abstracts{
We show how to obtain perfect samples from a quantum Gibbs state on a quantum computer. 
To do so, we adapt one of the ``Coupling from the Past''-algorithms 
proposed by Propp and Wilson. The algorithm has a probabilistic run-time
and produces perfect samples without any previous knowledge of the mixing time of a quantum Markov chain.
To implement it, we assume we are able to perform the phase estimation algorithm for the underlying 
Hamiltonian and implement a quantum Markov chain such that the transition probabilities between eigenstates only
depend on their energy.
We provide some examples of quantum Markov chains that satisfy these conditions 
and analyze the expected run-time of the algorithm, which depends strongly on the degeneracy of the underlying Hamiltonian.
For Hamiltonians with highly degenerate spectrum, it is efficient, 
as it is polylogarithmic in the dimension and linear in the mixing time.
For non-degenerate spectra, its runtime is essentially the same as its classical counterpart, which is linear in the mixing time and quadratic in the 
dimension, up to a logarithmic factor in the dimension.
We analyze the circuit depth necessary to implement it, which is proportional to the sum  of the depth necessary to implement one step of the quantum Markov chain and one 
phase estimation.
This algorithm is stable under noise in the implementation of different steps. We also briefly discuss how to adapt different ``Coupling from the Past''-algorithms to the quantum setting.
}{}{}

\vspace*{10pt}

\vspace*{3pt}

\vspace*{1pt}\textlineskip    

\section{Introduction}
Markov chain Monte Carlo methods are ubiquitous in science. 
They have a similar structure: the solution to a problem is encoded in the stationary distribution of a Markov chain that can
be simulated.
The chain is then simulated for a ``long enough'' time until the current state of the chain is ``close enough''
to a sample of the stationary distribution of interest.

It is expected that with the advent of quantum computers one could use similar methods to develop algorithms
to simulate quantum many-body systems that do not suffer from the sign problem~\cite{suzuki2012quantum}, and many quantum algorithms with this property were proposed
~\cite{terhalgibbs,poulin2009sampling,ge15rapid,yung2012quantum,riera2012thermalization,temme2011quantum,bilgin2010preparing,Kastoryano2016}.

However, as for classical Monte Carlo methods, it is in general difficult to obtain rigorous bounds on how long is ``long enough'', 
as the huge literature dedicated to Markov chain mixing attests \cite{markovmixing}. 
This prompted research on an algorithm that would ``decide for itself'' when the current state of the Markov chain is close to or is even a perfect sample of the stationary distribution,
without any prior knowledge of the mixing properties of the chain.

One of the first algorithms to do so was developed in \cite{lovasz1995exact}. 
Later Propp and Wilson proposed the ``Coupling from the Past'' (CFTP)-algorithm \cite{propp1996exact} and showed how it can be applied
to efficiently obtain perfect samples for the Ising model. 
They also showed how to sample perfectly from the stationary distribution of an unknown Markov chain for which 
we can only observe transitions
in a subsequent paper \cite{Propp1998170} 
and many perfect sampling algorithms were developed\footnote{The website \url{http://dimacs.rutgers.edu/~dbwilson/exact/}, maintained by Wilson,  contains a comprehensive list of references concerning the topic and other related material.}~ since. 
There are also proposals of quantum speedups to these algorithms \cite{PhysRevLett.104.250502}.

In this article, we will generalize some of these algorithms
to get perfect samples from a Gibbs state of a Hamiltonian on
a quantum computer. By this, we mean that we are able to perform any measurement and observe the same statistics for the outcomes as if we were measuring the actual 
Gibbs state.

To implement them, we will need to be able to perform the phase estimation algorithm~\cite{Kitaev2007} for the underlying Hamiltonian. 
Furthermore, we assume we can implement a quantum Markov chain
that drives the system to the desired Gibbs state and fulfills certain assumptions, such as reversibility of the chain, which we elaborate on below. 
We comment on which of the current 
proposals to prepare Gibbs states on quantum computers may be adapted for our purposes. 

Like it is the case for the classical algorithms,
our quantum algorithms do not require any previous knowledge about the mixing properties of the quantum Markov chain. They ``decide for themselves'' when the current state
of the system corresponds to a perfect sample of the target Gibbs state and their run-time is probabilistic.

We will focus on adapting the ``voter CFTP'' algorithm. 
For the classical voter CFTP
the expected run-time is quadratic in the dimension of the system and linear in the mixing time.
The run-time of our version will depend highly on the number of distinct eigenvalues of the Hamiltonian and the dimension of the eigenspaces.
In the worst case, which corresponds to Hamiltonians with a non-degenerate spectrum,
our version of this algorithm will turn out to have the same expected run-time as its classical counterpart, up to a logarithmic factor in the dimension. 
However, for Hamiltonians with degenerate spectra, our algorithm can be more efficient than the classical CFTP. In the case of Hamiltonians with an extremely degenerate spectrum,
our algorithm can even have a run-time which is proportional to the time necessary to obtain approximate samples.
We discuss how to explore this fact to obtain certifiably good samples efficiently for Hamiltonians whose spectrum can be ``lumped together'' into a small
number of intervals.
We also briefly discuss how to generalize other 
variations of CFTP.

These algorithms are stable under noise and we give bounds on their stability with respect to the implementation of the quantum Markov chain and the phase estimation steps.
A potential advantage of the ``voter CFTP'' in comparison to other methods proposed in the literature is that it only
requires the implementation of a quantum circuit of low depth a (potentially prohibitive) number of times and significant classical post-processing to obtain a perfect
sample without any prior knowledge of the mixing time.
Other methods require the implementation of a circuit of (potentially prohibitive) length one time to obtain an approximate sample, but only under the previous knowledge
of or assumptions on the mixing time. Therefore, these algorithms are qualitatively different from ours, as our  algorithm provides a certificate that we are obtaining good samples.

Another motivation for this work is to explore how coupling techniques for classical Markov chains may be applied and generalized to the quantum setting. These are one of the most 
useful tools to derive mixing times~\cite[Section 4.2]{markovmixing} and lie at the heart of many perfect sampling algorithms, as their name already suggests. 
The fact that we still lack a notion of a 
quantum analog of a coupling is, therefore, one of the main technical hurdles to generalize many results in the theory of classical Markov chains and is by itself an interesting open problem which 
hopefully this work can shed some light on.

\section{Preliminaries}
\subsection{Notation and basic concepts}
We begin by introducing some basic concepts we will need and fixing the notation. 
Throughout this paper, $\M_d$ will denote the space of $d\times d$ complex matrices and $[d]=\{1,\ldots,d\}$.
We will denote by $\D_d$ the set of $d$-dimensional quantum states, i.e. positive semidefinite matrices $\rho\in\M_d$ with trace $1$. 
We will call a Hermitian operator $H\in\M_d$ a Hamiltonian. We will always denote its spectral decomposition by $H=\sum_{i=1}^{d'}E_iP_i$, with $P_i$
orthogonal projections. Here $d'$ denotes the number of distinct eigenvalues of $H$.
As we will see later, the expected run-time of the algorithm will depend more on this number than the dimension.
The eigenspace corresponding to the energy level $E_i$ of $H$ will
be denoted by $S_i$ and we will denote its dimension by $|S_i|$.
When we write $H=\sum_{i=1}^{d}E_i\ketbra{\psi_i}$ we mean that $\{\ket{\psi_i}\}_{i=1}^d$ is an orthonormal eigenbasis of $H$.
For an inverse temperature $\beta>0$, we define 
$\mathcal{Z}_\beta=\tr[e^{-\beta H}]$ to be its partition function and $e^{-\beta H}/\mathcal{Z}_\beta$ its Gibbs state.
A quantum channel $T:\M_d\to\M_d$ is a trace preserving completely positive map. We will also refer to such a map as quantum Markov chain.
A state $\sigma\in\D_d$ is a stationary state of $T$ if we have $T(\sigma)=\sigma$.
The channel is called primitive if we have $\forall\rho\in\D_d:\lim\limits_{n\to\infty}T^n(\rho)=\sigma$ and $\sigma>0$.
There is an equivalent spectral characterization of primitive quantum channels. A quantum channel is primitive if 
$\sigma>0$
is the only eigenvector corresponding to eigenvalues of modulus $1$ of the channel~\cite{ergodicchiribella}. 
In particular, this implies that the property
of being primitive is stable under small perturbations of the channel.

A collection of self-adjoint operators $\{F_i\}_{i\in I}$ is called a POVM (positive-operator valued measure) if the $F_i\in\M_d$ are all
positive semidefinite and $\sum_{i\in I}F_i=\one$. Here $\one\in\M_d$ is the identity matrix. 
A state $\rho\in\D_d$ and a POVM  $\{F_i\}_{i\in I}$ induce a probability distribution $p$ through $p(i)=\tr[F_i\rho]$.
All the algorithms we will discuss have as their goal to produce exact samples of the distribution $p$ generated by
an arbitrary POVM in the case that $\rho$ is a Gibbs state.

The following class of quantum channels will be one of the backbones of the algorithms we will present later.

\hfill
\begin{defn}[Eigenbasis preserving quantum channels]
A quantum channel $T:\M_d\to\M_d$ is called eigenbasis preserving for a Hamiltonian $H=\sum\limits_{i=1}^{d'}E_iP_i$ and inverse temperature $\beta>0$
if we have that for all $i,j\in[d']$ the commutator
\[
[T(P_i),P_j]=0  
\]
vanishes and $T\lb\frac{e^{-\beta H}}{\mathcal{Z}_\beta}\rb=\frac{e^{-\beta H}}{\mathcal{Z}_\beta}$.

\end{defn}
\hfill\break 
By the commutator property, we can model the dynamics under $T$ on states that commute with $e^{-\beta H}/\mathcal{Z}_\beta$ as a classical Markov chain.  
One should not take this condition to imply that the dynamics under $T$ are classical, as will become clear in subsection \ref{sec:exchannels}, 
where we present some examples of eigenbasis preserving channels.
We will first suppose that we can implement these channels exactly, but will later relax this condition 
and discuss the influence of noise in section \ref{sec:noise}.

We will need some distinguishability measures for quantum states and channels and convergence speed measures
for primitive quantum channels.
One of the main ones is through the Schatten $1-$Norm $\|X\|_1=\tr[|X|]$ for $X\in\M_d$. 
This is justified by the clear operational interpretation given by its variational expression \cite[p. 404]{nielsen2000quantum}. If we 
denote by $\mathcal{P}_d$ the set of orthogonal projections in $\M_d$, we have for $\rho,\sigma\in\D_d$
\begin{align}\label{equ:variationall1}
\frac{\|\rho-\sigma\|_1}{2}=\sup\limits_{P\in\mathcal{P}_d}\tr[P(\rho-\sigma)]. 
\end{align}
That is, $\|\rho-\sigma\|_1/2$ expresses the maximal probability of correctly distinguishing two states $\sigma,\rho$ by a projective measurement.
This norm induces the $1\to1$ norm on operators $T:\M_d\to\M_d$:
\begin{align}
\|T\|_{1\to1}=\sup\limits_{X\in\M_d}\frac{\|T(X)\|_1}{\|X\|_1}. 
\end{align}

As a measure of the convergence speed of a quantum channel, we define the $l_1$-mixing time threshold of a primitive
quantum channel $T:\M_d\to\M_d$ with unique stationary state $\sigma$, which is given by
\begin{align*}
t_{\textrm{mix}}=\min\{n\in\N:\sup\limits_{\rho\in D_d}\|T^{n}(\rho)-\sigma\|_1\leq 2e^{-1}\}. 
\end{align*}
We will say that a channel $T:\M_d\to\M_d$ satisfies detailed balance or is reversible with respect to $\gibbs$ if we have that
\begin{align*}
T\lb\lb\gibbs\rb^{\frac{1}{2}} X\lb\gibbs\rb^{\frac{1}{2}}\rb=\lb\gibbs\rb^{\frac{1}{2}}T^*\lb X\rb \lb\gibbs\rb^{\frac{1}{2}} 
\end{align*}
holds for all $X\in\M_d$. Here $T^*$ is the adjoint of the channel with respect to the Hilbert-Schmidt scalar product. Satisfying detailed balance
with respect to $\gibbs$
implies in particular that $\gibbs$ is a stationary state of the channel.

A crucial ingredient for our sampling algorithm is the phase estimation algorithm, discovered originally in \cite{Kitaev2007}. 
There are now many variations of it \cite{abrams1999quantum,cleve1998quantum,Svore:2014:FPE:2600508.2600515} 
and it is still the subject of active research.
We will neither discuss in detail how to implement it nor its complexity and refer to \cite{Svore:2014:FPE:2600508.2600515} for that.
For our purposes, we will just suppose that for a given Hamiltonian $H$ acting on $\C^d$ we may implement a unitary 
$U$ on $\C^d\otimes(\C^2)^{\otimes t}$ for some $t\in\N$ that acts as follows:

For $\ket{\psi_i}$ an eigenstate of a Hamiltonian $H$ with $H\ket{\psi_i}=E_i\ket{\psi_i}$ we have $U\ket{\psi_i}\otimes\ket{0}=\ket{\psi_i}\otimes\ket{E_i}$, where $\ket{E_i}$ is the binary
expansion of $E_i$ in the computational basis of $(\C^2)^{\otimes t}$. 
We will first assume that we may implement $U$ exactly, but later discuss how imperfections in the implementation of the phase estimation
algorithm influence the output of the sampling algorithm in section \ref{sec:noise}.

We will now fix some notation and terminology for classical Markov chains.
A sequence $X_0,X_1,X_2,\ldots$ of random variables taking values in a (finite) set $S$, referred to as the state space,  is called a Markov chain if we have
\[
 P(X_{n+1}=j|X_n=i)=\pi(i,j)
\]
for a $|S|\times|S|$ matrix $\pi$. $\pi$ is called the transition matrix of the chain. 
We will always denote by $\pi$ the transition matrix of a Markov chain that should be clear from context.
Most of the times it will be the one induced by an eigenbasis-preserving channel through
\begin{align*}
\pi(i,j)=\tr\lb T(\ketbra{\psi_i})\ketbra{\psi_j}\rb. 
\end{align*}
A probability distribution $\mu$ on $S$ is called stationary if we have that $\pi\mu=\mu$. 
A Markov chain is said to be irreducible if
\[
 \forall i,j\in S ~\exists n:\pi^n(i,j)>0.
\]
It is aperiodic if 
\[
\forall i\in S: \text{gcd} \{n\in\N\setminus\{0\}:\pi^{n}(i,i)>0\}=1.
\]
Analogously to the quantum case, we say that the transition matrix $\pi$ satisfies detailed balance with respect to $\mu$ if
\begin{align*}
\pi(i,j)\mu(i)=\mu(j)\pi(j,i). 
\end{align*}
Satisfying detailed balance again implies that $\mu$ is stationary.
It is a well-known fact that if a Markov chain is aperiodic and irreducible there exists a unique stationary distribution $\mu$ such that for any other distribution $\nu$ on $S$ we have that 
$\lim\limits_{n\to\infty}\pi^n\nu=\mu$. We define the variational distance between probability distributions $\nu, \mu$ as

\begin{equation}\label{equ:deftvnorm}
\|\nu-\mu\|_1=\sum\limits_{i\in S}|\mu(i)-\nu(i)|. 
\end{equation}

With a slight abuse of notation, we will also denote the $l_1$-mixing time threshold in variation distance for a Markov chain by 
\begin{equation}\label{equ:deftmixclassical}
t_{\textrm{mix}}=\min\{n\in\N:\sup_\nu\|\pi^n\nu-\mu\|_1\leq 2e^{-1}\}.  
\end{equation}

Let 
\begin{equation*}
C=\min\{T|\forall i\in S\exists 1\leq k\leq T \text{ such that } X_k=i\}. 
\end{equation*}

We will denote by $E_i(C)$ the expected time it takes 
to observe all states starting from $X_0=i$ and by $T_C=\max_{i\in S}E_i(C)$ the cover time of the chain.
We refer to e.g. \cite[Chapter 1]{markovmixing} for a review of these basic concepts of Markov chains.  

\subsection{Lumpable Channels and Chains}
Given an equivalence relation or equivalently a partition of the state space $S=\bigsqcup_{i=1}^{d'}S_i$ of a Markov chain $X_0,X_1,X_2,\ldots$, 
it is possible to define a new stochastic process in which the state space is given
by the equivalence classes in the following way. 
Define the function $f:S\to[d']$ which maps a state to its equivalence class, that is
\begin{align*}
f(x)=i\Leftrightarrow x\in S_i. 
\end{align*}
This new stochastic process is then given by the random variables $f(X_0),(X_1),f(X_2),\ldots$.
If this stochastic process is again a Markov chain for all possible initial probability distributions on $S$, the chain is said to be lumpable with respect to this equivalence relation. 
We refer to~\cite[Section 6]{kemeny1960finite} for more on lumpable chains. These are sometimes also called projective chains.
The next Theorem gives necessary and sufficient conditions for lumpability.
\hfill\break 
\begin{thm}[Lumpable Chain]\label{thm:lumpablechains}
A necessary and sufficient condition for a Markov chain to be lumpable with respect to a partition $S=\bigsqcup_{i=1}^{d'}S_i$
is that for every pair $S_l,S_k$ we have for all $l,l'\in S_l$
\begin{align}\label{equ:transisiontslumped}
\sum\limits_{k\in S_k}\pi(l,k)=\sum\limits_{k\in S_k}\pi(l',k) 
\end{align}
Moreover, the transition probability between $S_l$ and $S_k$ in the lumpable chain is given by Eq. \eqref{equ:transisiontslumped}.
\end{thm}
\hfill\break 
\textbf{Proof:} We refer to \cite[Theorem 6.3.2]{kemeny1960finite} for a proof.
\qed
\hfill\break

In order to perfectly sample from Gibbs states with degenerate spectra, we will need the concept of a lumpable channel, which we introduce here.
\hfill\break
\begin{defn}[Lumpable channel]
An eigenbasis preserving quantum channel $T:\M_d\to\M_d$ is called lumpable for a Hamiltonian $H=\sum_{i=1}^{d'}E_iP_i$ at  inverse temperature $\beta>0$
if it is reversible and there is a function $f:\R\times\R\to[0,1]$ such that
\[
\tr\left(T(\ketbra{\psi_i})\ketbra{\psi_j}\right)=f(E_i,E_j)   
\]
for any unit vectors $\ket{\psi_i}\in P_i(\C^d)$ and $\ket{\psi_j}\in P_j(\C^d)$.
\end{defn}
\hfill\break 
Here $P_j(\C^d)$ denotes the image of the projection. 
That is, the transition probabilities for eigenstates of $H$ depend only on their respective energies.
Notice that as we demand that the quantum channel
satisfies detailed balance, the classical transition matrix induced by the channel will also satisfy detailed balance. 
The definition of lumpable channels is motivated by the following lemma.
\hfill\break 
\begin{lem}[Lumping energy levels]\label{lem:lumpedchain}
Let $T:\M_d\to\M_d$ be a lumpable quantum channel for a Hamiltonian $H=\sum_{i=1}^dE_i\ketbra{\psi_i}$ and inverse temperature $\beta>0$. 
Here $\{\ket{\psi_i}\}_{i=1}^{d}$ is an orthonormal eigenbasis of $H$.
Define the classical Markov chain
on $[d]$ with transition matrix given by
\begin{align*}
 \pi(i,j)=\tr(T(\ketbra{\psi_i})\ketbra{\psi_j})
\end{align*}
and partition the state space according to their energy into $\{S_1,\ldots,S_{d'}\}$, that is, the equivalence relation on the state space is given by
\begin{align}
 i\sim j\Leftrightarrow \tr(H\ketbra{\psi_i})=\tr(H\ketbra{\psi_j}).
\end{align}
Then the Markov chain is lumpable with respect to this partition and the lumped chain has transition matrix
\begin{align}\label{equ:transitionlumpable}
\tilde{\pi}(S_l,S_k)=|S_k|f(E_l,E_k). 
\end{align}
Moreover, the stationary distribution of the lumped chain is given by
\begin{align}
\tilde{\mu}(S_i)=|S_i|\frac{e^{-\beta E_i}}{\mathcal{Z}_\beta},
\end{align}
where $|S_i|$ is the degeneracy of the energy level, and the chain satisfies detailed balance with respect to $\tilde{\mu}$.
\end{lem}
\hfill\break 
\noindent\ignorespacesafterend
\textbf{Proof:} It follows from Theorem \ref{thm:lumpablechains} that it is sufficient and necessary for the chain to be
lumpable that 
for $S_l,S_k$ we have for all $l,l'\in S_l$
\begin{align}\label{equ:conditionlumpable}
\sum\limits_{k\in S_k}\pi(l,k)=\sum\limits_{k\in S_k}\pi(l',k). 
\end{align}
Eq. \eqref{equ:conditionlumpable} holds for lumpable quantum channels, as we have $\sum\limits_{k\in S_k}\pi(l,k)=|S_k|f(E_l,E_k)$, which 
clearly only depends on the equivalence class of $l$.
Therefore, we may define a Markov chain with respect to this partition and from Theorem \ref{thm:lumpablechains} it
follows that the transition matrix of the lumpable chain is given by \eqref{equ:transitionlumpable}. We will now show that it satisfies detailed balance with 
respect to $\tilde{\mu}$ and therefore $\tilde{\mu}$ is the stationary distribution of the chain. We have
\begin{align}\label{equ:beforedetailedbalance}
\tilde{\mu}(S_i)\tilde{\pi}(S_i,S_j)= |S_i|\frac{e^{-\beta E_i}}{\mathcal{Z}_\beta}|S_j|f(E_i,E_j). 
\end{align}
Now, as the original chain satisfied detailed balance, it holds that
\begin{align}\label{equ:detailebalancechain}
\frac{e^{-\beta E_i}}{\mathcal{Z}_\beta}f(E_i,E_j)=f(E_j,E_i)\frac{e^{-\beta E_j}}{\mathcal{Z}_\beta}. 
\end{align}
Plugging Eq. \eqref{equ:detailebalancechain} into \eqref{equ:beforedetailedbalance} we see that the lumpable chain 
satisfies detailed balance with respect to $\tilde{\mu}$. This implies that $\tilde{\mu}$ is the stationary distribution of the lumped Markov chain.
\qed
\hfill\break

When working with lumpable channels, $t_{\textrm{mix}}$ will always refer to the mixing time of the lumped chain.
It is in general not clear how the mixing time of the lumpable chain relates to the mixing time of the original chain and this is a topic of current research. Surprisingly, the mixing time may even increase under lumping,
as was shown recently in~\cite{sensitivitymixing}. However, as is shown e.g. in~\cite[Lemma 12.8]{markovmixing} and remarked in~\cite{sensitivitymixing}, important parameters that describe the 
convergence of a Markov chain, such as the spectral gap or Cheeger constant~\cite[Chapter 12]{markovmixing}, can only increase when lumping chains. This implies that the mixing time cannot increase significantly
by lumping. For example, in the counterexample found in~\cite{sensitivitymixing}, lumping increases the mixing time by a factor of $\Theta(\log(d))$. 

\subsection{(Classical) Voter CFTP}\label{sec:cftpalgos}

We will briefly describe a perfect sampling algorithms based on CFTP for Markov chains introduced in  \cite{Propp1998170}, called ``voter CFTP''. 
We mostly stick to their terminology and notation.
The goal of this algorithm is to produce perfect samples of the stationary distribution $\mu$ of some Markov chain. One of the main advantages of this algorithm is that we only
need to be able to observe valid transitions of this Markov chain to obtain perfect samples of the target distribution.

One should note that this is in general not the most efficient algorithm for perfect sampling \cite{Propp1998170}, but arguably the simplest to understand. 
Besides the pedagogical motivation to present it, it turns out that this version is of interest in the quantum case, as we will see later.
For this algorithm we suppose we have access to a randomized procedure $\mathbf{RandomSuccessor}:S\to S$ such that  $P(\mathbf{RandomSuccessor}(i)=j)=\pi(i,j)$, where $\pi$ is a transition matrix having $\mu$ 
as a stationary measure.
Let $G$ be a vertex-labeled graph with vertices $-\N_0\times S$ and labels $S$. We will define the labels and edges as the algorithm runs and denote by $G(k,i)$ the label of the vertex $(k,i)$. 
Pseudocode for the algorithm is provided 
below in algorithm \ref{algo:cftppropp} .

\afterpage{
\begin{algorithm}[tp!]

\begin{algorithmic}[1]
\Procedure{Voter CFTP}{}
\State Set $G(0,i)=i$ and $k=0$.
\While {$\exists j,i\in S$ s.t. $G(k,i)\not=G(k,j)$}
\For {$i\in$ $S$}
\State Let $j=\mathbf{RandomSuccessor}(i)$. 
\State Set $G(k-1,i)=G(k,j)$.
\State Add the edge $\{(k-1,i),(k,j)\}$\label{step:addedgeclassical}
\EndFor
\State Set $k\to k-1$
\EndWhile
\\
\Return $G(k,i_0)$ for some $i_0\in S$
\EndProcedure
\end{algorithmic}
\vspace*{13pt}

\fcaption{Voter CFTP~\cite{Propp1998170}}\label{algo:cftppropp}

\end{algorithm}

\begin{figure}[bp!]\label{fig:relationineqcla}
\centering

\begin{tikzpicture}
\node[draw] (r1c1) at (5,0) {1};
\node[draw] (r2c1) [below =of r1c1] {2};
\node[draw] (r3c1) [below =of r2c1] {3};

\node[draw] (r1c2) [left =of r1c1] {2};
\node[draw] (r2c2) [below =of r1c2] {3};
\node[draw] (r3c2) [below =of r2c2] {1};

\node[draw] (r1c3) [left =of r1c2] {0};
\node[draw] (r2c3) [below =of r1c3] {0};
\node[draw] (r3c3) [below =of r2c3] {0};

\draw[-,] (r1c2) to (r2c1);
\draw[-,] (r2c2) to (r3c1);
\draw[-,] (r3c2) to (r1c1);

\end{tikzpicture} 
\vspace*{13pt}
\fcaption{Possible first two columns of the graph after running the for-loop in the fourth step one time for $d=3$. Notice that the third column has still not
been labeled.} 
\end{figure}

\begin{figure}[bp!]\label{fig:relationineqcla1}
\centering
\begin{tikzpicture}
\node[draw] (r1c1) at (5,0) {1};
\node[draw] (r2c1) [below =of r1c1] {2};
\node[draw] (r3c1) [below =of r2c1] {3};

\node[draw] (r1c2) [left =of r1c1] {2};
\node[draw] (r2c2) [below =of r1c2] {3};
\node[draw] (r3c2) [below =of r2c2] {1};

\node[draw] (r1c3) [left =of r1c2] {3};
\node[draw] (r2c3) [below =of r1c3] {3};
\node[draw] (r3c3) [below =of r2c3] {3};

\draw[-,] (r1c2) to (r2c1);
\draw[-,] (r2c2) to (r3c1);
\draw[-,] (r3c2) to (r1c1);
\draw[-,] (r1c3) to (r2c2);
\draw[-,] (r2c3) to (r2c2);
\draw[-,] (r3c3) to (r2c2);

\end{tikzpicture} 
\vspace*{13pt}
\fcaption{
Possible graph after running the for-loop one more time. Notice the algorithm has terminated and outputs the sample $3$. 
}

\end{figure}

\clearpage
}
One does not need to add the edge in step \ref{step:addedgeclassical}. This only helps to visualize the process.
The expected run-time of this algorithm and its complexity, of course, depend on properties of $\mathbf{RandomSuccessor}:S\to S$.
We will discuss these when we analyze the same questions for our algorithm in the quantum case.
We now provide a proof that algorithm \ref{algo:cftppropp} indeed produces a perfect sample if it terminates almost surely.
\hfill\break 
\begin{thm}
Suppose algorithm \ref{algo:cftppropp} terminates with probability $1$ and denote the output by $Y$.
Then $P(Y=i)=\mu(i)$.
\end{thm}
\hfill\break 
\noindent\ignorespacesafterend
\textbf{Proof:} Let $\epsilon>0$. As the algorithm terminates with probability $1$, there is a $N_\epsilon$ such that
\begin{align*}
 P(\text{algorithm terminates after at most } N_\epsilon \text{ steps})\geq1-\epsilon. 
\end{align*}
Denote by $A_\epsilon$ the event that the algorithm terminates after at most $N_\epsilon$ steps. Define a Markov chain 
$X_{-M},X_{-M+1},X_{-M+2},\ldots,X_0$ for some $M\in \N$ and choose $X_{-M}$ according to $\mu$, i.e. $P(X_{-M}=i)=\mu(i)$.
The transitions are defined by the graph, which we suppose was labeled
for all $(k,i)$ with $k>-M$. 
Given $X_{k}=j$ we set $X_{k+1}=i$, where $\{(k,j),(k+1,i)\}$ is an edge of the graph $G$.
As we chose $X_{-M}$ according to $\mu$ and $\mathbf{RandomSuccessor}$ has $\mu$ as a stationary distribution, $P(X_{k}=i)=\mu(i)$ for all 
$-M\leq k\leq0$.
We have
\[
P(X_0\not=Y)=P(X_0\not=Y|A_\epsilon)P(A_\epsilon)+P(X_0\not=Y|A_\epsilon^C)P(A_\epsilon^C). 
\]
One can check that the label on the graph at $(-M,i)$ is nothing but the value of $X_0$. Thus, if we assume that the algorithm
has terminated, the value of $X_0$ does not depend on the initial value and will always be equal to $Y$. Therefore $P(X_0\not=Y|A_\epsilon)=0$ if $-M\leq N_\epsilon$. Also, by construction, 
$P(X_0\not=Y|A_\epsilon^C)P(A_\epsilon^C)\leq\epsilon$. We then conclude $P(X_0\not=Y)\leq\epsilon$ and so the value of $Y$ and $X_0$ coincide, as $\epsilon$ was arbitrary.
As $X_0$ is distributed according to $\mu$, so is $Y$.\qed
\hfill\break

\section{CFTP for quantum Gibbs states}
\subsection{Voter CFTP}
We will now show how to adapt the voter CFTP algorithm to quantum Gibbs state. We start by focusing on Hamiltonians that have a non-degenerate spectrum, as we need less assumptions in this case and the  proof is simpler.
We later generalize to arbitrary Hamiltonians.
Given a Hamiltonian $H\in\M_d$ with non-degenerate spectrum, an eigenbasis preserving quantum channel $T$ for some inverse temperature $\beta>0$ and a POVM $\mathcal{F}=\{F_i\}_{i\in I}$, the following
algorithm allows us to obtain perfect samples from the distribution $p(i)=\tr\left[F_i\frac{e^{-\beta H}}{\mathcal{Z}_\beta}\right]$. 
The algorithm uses three registers corresponding to the tensor factors $\C^{d}\otimes\lb\C^2\rb^{\otimes t}\otimes\lb\C^2\rb^{\otimes t}$, where $t$ is large enough to perform phase estimation for $H$ and tell apart the different eigenvalues of $H$.
We will discuss how to choose $t$ in section \ref{sec:noise}.
The first one will encode the current state of our system, while the other two will be used to record the output of two phase estimation steps.
Define a labeled graph $G$ with vertices $V=-\N_0\times\{1,\ldots,d\}$ and labels given by $\{0,\ldots,d\}$. We assume that $G$ has no edges at the beginning of the algorithm and the vertices are labeled as
\begin{equation}\label{equ:defngraph}
G(k,j)=
\begin{cases}
j &\mbox{if } k=0
\\ 0 &\mbox{otherwise}
\end{cases} . 
\end{equation}

We assume we can prepare the maximally mixed state $\frac{\one}{d}$. 
This can be done by picking a uniformly distributed integer between $1$ and $d$ and preparing the corresponding state
of the computational basis, for example. We will assume that the Hamiltonian has a spectral decomposition given by $H=\sum_{i=1}^dE_i\ketbra{\psi_i}$. 
The number $n$ denotes how many samples we wish to obtain in total and $c$ will denote a counter for the number of samples we still wish to obtain.
The pseudocode for the algorithm is below in algorithm \ref{algo:cftp}.

\begin{algorithm}
\vspace*{13pt}

\begin{algorithmic}[1]
\Procedure{Quantum Voter CFTP (non-degenerate case)}{}
\State Set $R=\varnothing$ and $c=n$.
\While {$c\not=0$}\label{step:whileloop}

\State Prepare the state $\frac{\one}{d}\otimes\ketbra{0}\otimes\ketbra{0}$
\State Run phase estimation on the first and second register.
\State Measure the second register in the computational basis. \label{step:firstmeasurephase}
\If {$i\in R$}
\State Measure $\mathcal{F}$ on the first register. 
\State Update $c$ to $c-1$.
\Else
\State Apply $T\otimes\id_{2^t}\otimes\id_{2^t}$ to the system.
\State Run phase estimation on the first register and third register.
\State Measure the third register in the computational basis. Let the result be $j$.\label{step:secondmeasurephase}
\State For the largest $k$ s.t. $G(k,i)=0$ we add the edge $\{(k,i),(k+1,j)\}$.\label{step:largestfreek}
\If {$G(k+1,j)\not=0$} 
\State Change the labels on all the vertices $(k',i')$ with $k'<k$ for which there is a path to $(k,i)$ from $0$ to $G(k+1,j)$.\label{step:labelgraph}
\EndIf
\If{There is $k_0\in-\N$ and $l\in[d]$ s.t. $\forall i\in[d]$ $G(k_0,i)=l$} \label{step:checkingconst}
\State Append $l$ to $R$.
\State Erase all edges to the vertices $(k_0,i)$ and set the labels to $G(k_0,i)=i$ and $G(k,i)=0$ for $k<k_0$.
\EndIf
\EndIf
\EndWhile
\EndProcedure
\end{algorithmic}
\fcaption{Voter CFTP for quantum Gibbs states}\label{algo:cftp}
\end{algorithm}

We now prove it indeed outputs perfect samples.
\hfill\break 
\begin{thm}\label{thm:algoconverges}
 Let $T$ be a primitive, eigenbasis preserving quantum channel for a Hamiltonian $H$ and inverse temperature $\beta>0$. 
 Then algorithm \ref{algo:cftp} terminates with probability 1 and generates $n$ perfect samples of the distribution $p$ defined above.
\end{thm}
\hfill\break 
\noindent\ignorespacesafterend
\textbf{Proof:} We will first show that with probability $1$ there is a $k\in-\N$ and $l\in[d]$ such that $\forall i\in[d]$ $G(k,i)=l$.
The probability that we observe  an eigenstate $\ket{\psi_i}$ at step \ref{step:firstmeasurephase} is $\frac{1}{d}$, so with probability 1 we will observe it if we run the loop at step \ref{step:whileloop} often enough. 
This implies that we will assign a label different to $0$ to arbitrary vertices of the graph $G$ if we run the while-loop at step \ref{step:whileloop} for long enough.
Observe that as $T$ is an eigenbasis preserving quantum channel, the dynamics on the eigenbasis of $H$ under $T$ is just a classical Markov chain. 
As $T$ is primitive and the stationary state has full rank, 
this Markov chain is aperiodic and irreducible \cite{ergodicchiribella}. 
Because of that, the probability
that we will obtain a $k$ such that $G(k,i)=l$ $\forall i\in[d]$ is $1$, using the same argument as the one given in \cite{Propp1998170} for the classical case.
By the same argument, the probability that this label is $l$ is given by $\frac{e^{-\beta E_l}}{\mathcal{Z}_\beta}$, 
as this is the stationary distribution of the underlying classical Markov chain.
As before, we will observe $\ket{\psi_l}\ket{E_l}\ket{0}$ at step \ref{step:firstmeasurephase} with probability 1 if we run the while-loop at step \ref{step:whileloop} often enough 
and this will then be a perfect sample by the previous discussion.
\qed
\hfill\break

Note that we could check if the measurement outcome we observe at step \ref{step:secondmeasurephase} is one of the desired outcomes to increase our probability of observing it.

\begin{figure}\label{fig:relationineq}
\centering

\begin{tikzpicture}
\node[draw] (r1c1) at (5,0) {1};
\node[draw] (r2c1) [below =of r1c1] {2};
\node[draw] (r3c1) [below =of r2c1] {3};

\node[draw] (r1c2) [left =of r1c1] {2};
\node[draw] (r2c2) [below =of r1c2] {3};
\node[draw] (r3c2) [below =of r2c2] {0};

\node[draw] (r1c3) [left =of r1c2] {2};
\node[draw] (r2c3) [below =of r1c3] {2};
\node[draw] (r3c3) [below =of r2c3] {0};

\node[draw] (r1c4) [left =of r1c3] {2};
\node[draw] (r2c4) [below =of r1c4] {0};
\node[draw] (r3c4) [below =of r2c4] {0};

\draw[-,] (r1c2) to (r2c1);
\draw[-,] (r2c2) to (r3c1);
\draw[-,] (r2c3) to (r1c2);
\draw[-,] (r1c3) to (r1c2);
\draw[-,] (r1c4) to (r2c3);

\end{tikzpicture} 
\vspace*{13pt}

\fcaption{Possible first four columns of the graph after running the while-loop in step \ref{step:whileloop} five times for $d=3$.} 
\end{figure}

\begin{figure}\label{fig:relationineq1}
\centering
\begin{tikzpicture}
\node[draw] (r1c1) at (5,0) {1};
\node[draw] (r2c1) [below =of r1c1] {2};
\node[draw] (r3c1) [below =of r2c1] {3};

\node[draw] (r1c2) [left =of r1c1] {2};
\node[draw] (r2c2) [below =of r1c2] {3};
\node[draw] (r3c2) [below =of r2c2] {3};

\node[draw] (r1c3) [left =of r1c2] {2};
\node[draw] (r2c3) [below =of r1c3] {2};
\node[draw] (r3c3) [below =of r2c3] {2};

\node[draw] (r1c4) [left =of r1c3] {2};
\node[draw] (r2c4) [below =of r1c4] {0};
\node[draw] (r3c4) [below =of r2c4] {0};

\draw[-,] (r1c2) to (r2c1);
\draw[-,] (r2c2) to (r3c1);
\draw[-,] (r2c3) to (r1c2);
\draw[-,] (r1c3) to (r1c2);
\draw[-,] (r1c4) to (r2c3);
\draw[-,] (r3c2) to (r3c1);
\draw[-,] (r3c3) to (r1c2);

\end{tikzpicture} 
\vspace*{13pt}
\fcaption{
Possible graph after running the while-loop two more times. Notice the algorithm has terminated and outputs the sample $2$. 
}

\end{figure}

Given how many distinct eigenvalues the Hamiltonian has and that we are able to implement a lumpable channel,
we may run a modified version of algorithm \ref{algo:cftp} and obtain perfect samples.
The steps of the algorithm are exactly the same and we do not write them out in detail. 
The only difference is the graph we feed the transitions to and what we feed.
Let $d'$ be again the number of distinct eigenvalues of $H$. In the case of degenerate Hamiltonians, we define 
a labeled graph $G$ with vertices $V=-\N_0\times\{1,\ldots,d'\}$ and labels given by $\{0,\ldots,d'\}$. We assume that $G$ has no edges at the beginning of the algorithm and the vertices are labeled as
\begin{equation}\label{equ:defngraph2}
G(k,j)=
\begin{cases}
j &\mbox{if } k=0
\\ 0 &\mbox{otherwise}
\end{cases} . 
\end{equation}
That is, the graph is essentially the same as in the non-degenerate case but with $d'$ instead of $d$ labels and vertices.
At step \ref{step:largestfreek} we then label the graph according to the energy levels we measured before at steps \ref{step:firstmeasurephase} and \ref{step:secondmeasurephase}, as we can only tell apart states with different energies
using phase estimation. We then have:
\hfill\break 
\begin{thm}\label{thm:lumpingalgorithm}
Let $T$ be a primitive, lumpable quantum channel for a Hamiltonian $H$ and inverse temperature $\beta>0$. 
Suppose further that $H$ has $d'$ distinct eigenvalues.
Then, if we run algorithm \ref{algo:cftp} with a graph modified as explained above, it terminates with probability 1 and generates $n$ perfect samples of the distribution $p(i)=\tr\left(F_i\frac{e^{-\beta H}}{\mathcal{Z}_\beta}\right)$.
\end{thm} 

\hfill\break 
\noindent\ignorespacesafterend
\textbf{Proof:} It should be clear that in this case the classical CFTP algorithm we are running based on the measurement outcomes
will generate perfect samples from the stationary distribution of the lumped chain defined in Lemma \ref{lem:lumpedchain}. The convergence
is guaranteed by the same argument as in the proof of Theorem \ref{thm:algoconverges}. From Lemma \ref{lem:lumpedchain} it follows that we will obtain the sample $S_j$ with probability 
\begin{align}
|S_j|\frac{e^{-\beta E_j}}{\mathcal{Z}_\beta}. 
\end{align}
Now, given that we have observed the label associated to $S_j$ after the first phase estimation step, we know that the state of the first register is
given by
\begin{align}
\rho_j=\frac{1}{|S_j|}P_j.
\end{align}
Measuring $F_i$ on the outputs of the algorithm, therefore, gives perfect samples from the distribution $p$.
\qed
\hfill\break

\subsection{Examples of eigenbasis preserving and lumpable channels}\label{sec:exchannels}
In order to run algorithm \ref{algo:cftp}, we need to be able to implement a primitive eigenbasis preserving quantum channel for the Gibbs state we want to sample from in the case of non-degenerate spectrum
and further that it is lumpable for the general case.
In recent years many algorithms have been proposed to approximately prepare quantum Gibbs states on a quantum computer~\cite{terhalgibbs,poulin2009sampling,ge15rapid,yung2012quantum,riera2012thermalization,temme2011quantum,bilgin2010preparing,Kastoryano2016}. 
We will here briefly discuss how some of them provide us with eigenbasis preserving or lumpable quantum channels for Gibbs states.

One class of eigenbasis preserving channels in the non-degenerate case are quantum dynamical semigroups with Davies generators. 
These are Markovian approximations for a quantum 
system weakly coupled to a thermal reservoir. A detailed description of the derivation and structure of Davies generators is beyond the scope of this article and can be found in \cite{dumcke1979proper,davies1976quantum}.
Under some conditions on the Hamiltonian and the coupling of the system to the bath, the Davies semigroup is primitive. 
The exact speed of this convergence is the subject of current research. We refer to \cite{lowerbounddavies}
for a discussion of the conditions under which the Davies generators are primitive and some bounds on the convergence speed. 
In \cite{Kastoryano2016} Davies generators are proposed as a way of preparing thermal states on a quantum computer.

For our purposes, their main relevant property is that 
if the underlying Hamiltonian has a non-degenerate spectrum, the
dynamics in the eigenbasis of the Hamiltonian does not couple diagonal terms to off-diagonal terms. They are therefore eigenbasis preserving.
This was observed by many authors since the beginning of their study \cite{davies1976quantum,davies1979generators,daviesqubits} and we refer to those for a proof of this claim.

More generally, it can be shown that dynamical semigroups that satisfy a quantum version of the detailed balance 
condition and whose stationary state has a non-degenerate spectrum are always eigenbasis preserving \cite{ALICKI1976249}.
Our two previous examples fall into that category.
This gives us a simple sufficient criterion to check whether a given implementation is eigenbasis preserving.

Note, however, that it is not a priori clear that a quantum dynamical semigroup can be implemented efficiently
or by only using local operations. We refer to \cite{Kastoryano2016,kliesch2011dissipative} for a discussion of these topics.

An example of a lumpable channel is given by the implementation of the quantum Metropolis algorithm proposed in \cite{temme2011quantum}, as the quantum channel implemented 
at each step maps eigenstates of $H$ to eigenstates of $H$ and the transition probabilities are a function of their energy difference.
However, it can be simplified for our purposes. 
As in the usual Metropolis algorithm, at each step we have to accept or reject a move that was made. 
One of the main difficulties to implement the quantum algorithm is reversing
the evolution of the system if we reject the move. This is because, by the No-Cloning Theorem \cite{wootters1982single}, 
we can't make a copy of the previous state of the system.
But the information that we rejected the move is enough for our algorithm, as we may simply copy the previous label 
when labeling
the vertices. Therefore, we may skip the procedure of reversing the move.

\subsection{Lumping Eigenstates together to obtain good samples}\label{sec:lumpingeigenvalues}
Until now we assumed we are able to implement phase estimation exactly and know the number of distinct eigenvalues of $H$.
We may loosen this assumption and lump different eigenvalues together.

\hfill 
\begin{defn}[$\epsilon$-Spectral Covering]
Let $H\in\M_d$ be a Hamiltonian and $\epsilon>0$ be given. We call 
$\{e_1,\ldots,e_{d'}\}\subset\R$
a $\epsilon$-Spectral Partition for $H$ if 
\begin{align*}
\sigma(H)\subset\bigsqcup_{i=1}^{d'}(e_i-\epsilon,e_i+\epsilon)
\end{align*}
and for all $i\in[d']:\sigma(H)\cap(e_i-\epsilon,e_i+\epsilon)\not=\varnothing$.
Here $\sigma(H)$ denotes the spectrum of $H$.
We will refer to $d'$ as the size of the covering.
\end{defn}
\hfill\break 
It should be clear that $\epsilon$-spectral coverings are not unique and may have different sizes for fixed $\epsilon$.
Although an $\epsilon$-spectral covering will not be readily available in most cases, there are some methods to obtain them. One can use e.g. the Gershgorin circle Theorem~\cite[Section VIII]{bhatia1997matrix} to obtain a covering.
If we can decompose $H$ into local commuting terms it is also possible to obtain an $\epsilon$-spectral covering by considering that the spectrum of $H$ must consist of
sums of the eigenvalues of the local terms.
Spectral coverings will be useful later to quantify the stability of algorithm \ref{algo:cftp} with respect to measuring the wrong energy with phase estimation. 
Here we will focus on showing how we may  still obtain good samples based on
an $\epsilon$-spectral covering and that the algorithm is stable w.r.t.  introducing degeneracies into the spectrum because we can only obtain an estimate of it to a finite precision.
Given an $\epsilon$-spectral covering for the Hamiltonian, we may run algorithm \ref{algo:cftp} with the number of labels
being given by the size of the covering. 
If we use the Metropolis algorithm from~\cite{temme2011quantum} with the $e_i$ as the possible energies to define the transition probabilities we obtain:
\hfill\break 
\begin{thm}\label{thm:lumpingesigenstates}
Let $H\in\M_d$ be a Hamiltonian and $\{e_1,\ldots,e_{d'}\}\subset\R$ be an $\epsilon$-spectral covering.
Suppose we run algorithm \ref{algo:cftp} with this $\epsilon$-spectral covering as described above. 
Then the  probability distribution $\tilde{p}$ of samples obtained from outputs of algorithm \ref{algo:cftp}
satisfies
\begin{align}\label{equ:tracefaultyphase}
\|\tilde{p}-p\|_1\leq\sqrt{4\epsilon\beta} 
\end{align}
\end{thm}
\hfill\break 
\noindent\ignorespacesafterend
\textbf{Proof:} The stationary distribution of the lumped chain will be 
\begin{align*}
\tilde{\mu}(i)=|\sigma(H)\cap(e_i-\epsilon,e_i+\epsilon)|\frac{e^{-\beta e_i}}{\tilde{\mathcal{Z}}_\beta},  
\end{align*}
with $\tilde{\mathcal{Z}}_\beta=\sum_{i=1}^{d'}|\sigma(H)\cap(e_i-\epsilon,e_i+\epsilon)|e^{-\beta e_i}$.
Let $\tilde{P}_i$ be the projection onto the subspace spanned by the eigenvectors of $H$ corresponding to eigenvalues in $\sigma(H)\cap(e_i-\epsilon,e_i+\epsilon)$.
From the proof of Theorem~\ref{thm:lumpingalgorithm}, it follows that algorithm \ref{algo:cftp} will output the state
\begin{align}
\tilde{\rho}=\frac{1}{\tilde{\mathcal{Z}}_\beta}\sum\limits_{i=1}^{d'}\frac{e^{-\beta e_i}}{|\sigma(H)\cap(e_i-\epsilon,e_i+\epsilon)|}\tilde{P}_i. 
\end{align}
We will now show $\left\|\tilde{\rho}-\frac{e^{-\beta H}}{\mathcal{Z}_\beta}\right\|_1\leq\sqrt{4\epsilon\beta}$,
from which the claim again follows from the variational definition of the trace norm.
From Pinsker's inequality~\cite[Theorem 3.1]{hiai1981}, it follows that
\begin{align}\label{equ:pinsker}
\left\|\tilde{\rho}-\frac{e^{-\beta H}}{\mathcal{Z}_\beta}\right\|_1\leq\sqrt{2D\lb\frac{e^{-\beta H}}{\mathcal{Z}_\beta}||\tilde{\rho}\rb}, 
\end{align}
where
$D\lb\frac{e^{-\beta H}}{\mathcal{Z}_\beta}||\tilde{\rho}\rb=\tr\lb\frac{e^{-\beta H}}{\mathcal{Z}_\beta}(\log(\frac{e^{-\beta H}}{\mathcal{Z}_\beta})-\log(\tilde{\rho})\rb$
is the relative entropy.
As $\tilde{\rho}$ and 
\begin{align*}
\frac{e^{-\beta H}}{\mathcal{Z}_\beta}=\frac{1}{\tr\lb e^{-\beta H}\rb}\sum\limits_{E_j\in\sigma(H)}e^{-\beta E_j}P_j 
\end{align*}
commute, we have
\begin{align}
D\lb\frac{e^{-\beta H}}{\mathcal{Z}_\beta}||\tilde{\rho}\rb=\sum\limits_{i=1}^{d'}\sum\limits_{E_j\in\sigma(H)\cap(e_i-\epsilon,e_i+\epsilon)}\frac{e^{-\beta E_j}}{\mathcal{Z}_\beta}\lb\log\lb\frac{\tilde{\mathcal{Z}}_\beta}{\mathcal{Z}_\beta}\rb
+\beta\lb E_j-e_i\rb\rb.
\end{align}
As we have an $\epsilon$-spectral covering, 
we have $E_j-e_i\leq\epsilon$ for $E_j\in\sigma(H)\cap(e_i-\epsilon,e_i+\epsilon)$  and $\frac{\tilde{\mathcal{Z}}_\beta}{\mathcal{Z}_\beta}\leq e^{\beta\epsilon}$. From this we obtain
\begin{align}\label{equ:boundrelativeentropy}
D\lb\frac{e^{-\beta H}}{\mathcal{Z}_\beta}||\tilde{\rho}\rb\leq2\beta\epsilon. 
\end{align}
Plugging Eq. \eqref{equ:boundrelativeentropy} into \eqref{equ:pinsker} we obtain the claim.
\qed
\hfill\break 

This result may be interpreted as a first stability result. 
This shows that if we lump eigenvalues that are very close together, the Gibbs state does not change a lot.
That is, if we introduce artificial degeneracies by not being able to tell apart eigenvalues that are very close through phase estimation this
will not change the output of the algorithm significantly.
As observed in \cite{temme2011quantum}, one could argue that a similar effect could in principle also affect classical
Markov chain methods, as we are only able to compute the transition probabilities up to a finite precision. This does not seem to
affect them in practice. 
Moreover, if we want samples that are certifiably at most $\delta$ apart in total variation distance at inverse temperature $\beta>0$,
we may lump together eigenvalues that are at most $\delta^2/4\beta$ apart. As we will see later, high levels of degeneracy can reduce the run-time of the algorithm
and this can be used to obtain good samples more efficiently.

\section{Stability of the Algorithm}
We will now address two possible sources of noise for algorithm \ref{algo:cftp} and show it is stable under these two. 
First, in the implementation of the channel and second in the phase estimation steps.

\subsection{Stability in the implementation of the Channel}\label{sec:noise}
As shown in \cite{perturbstat}, one may quantify the stability of primitive quantum Markov chains with the following constant:

\hfill 
\begin{defn}
 Let $T:\M_d\to\M_d$ be a primitive quantum channel with stationary state $\sigma\in\D_d$. We define 
\[
 \kappa(T)=\sup\limits_{X\in\M_d,\tr(X)=0}\frac{\|(\id-T+T_\infty)^{-1}(X)\|_1}{\|X\|_1}
\]
with $T_\infty(X)=\tr(X)\sigma$.
\end{defn}
\hfill\break 
We refer to \cite{perturbstat} for bounds on it and how it can be used to quantify the stability of a quantum Markov chain with respect 
to different perturbations.
Note that due to the spectral characterization of primitive quantum channels \cite{spectralconv}, 
the set of primitive quantum channels is relatively open in the convex set of quantum channels.
\hfill\break 
\begin{thm}\label{thm:stabilitychannel}
 Let $T:\M_d\to\M_d$ be a primitive eigenbasis preserving channel for a Hamiltonian $H$ and inverse temperature $\beta>0$ and $T':\M_d\to\M_d$ a quantum channel
 satisfying
\begin{equation}\label{equ:channelsepsclose}
\|T-T'\|_{1\to1}\leq \epsilon 
\end{equation}
for some $\epsilon>0$ small enough for $T'$ to be primitive too. 
For a POVM $\{F_i\}_{i\in I}$, let $p$ and $p'$ be probability distributions we obtain by measuring  $\{F_i\}_{i\in I}$ on the output of algorithm
\ref{algo:cftp} using $T$ and $T'$ respectively. Then
\begin{equation}\label{equ:probepsclose}
\|p-p'\|_1\leq(\kappa(T)+2)\epsilon.
\end{equation}

\end{thm}

\hfill\break 
\noindent\ignorespacesafterend
\textbf{Proof:} Let $\{P_i\}_{1\leq i\leq d'}$ be the eigenprojections of $H$ and define $Q:\M_d\to\M_d$ to be the quantum channel given by
\begin{equation}
 Q(X)=\sum\limits_{i=1}^{d'}\tr(P_iX)\frac{P_i}{|S_i|}.
\end{equation}
Note that as $T$ is an eigenbasis preserving channel, $QTQ$ is an eigenbasis preserving channel with stationary state $\frac{e^{-\beta H}}{\mathcal{Z}_\beta}$. 
As $T'$ is assumed to be primitive, $QT'Q$ is primitive too, as $\|QT'Q-QTQ\|_{1\to1}\leq\|T'-T\|_{1\to1}$.
Denote by $\rho$
the stationary state of the channel $QT'Q$.
By the variational expression for the trace distance, we have that
\begin{align}
\|p-p'\|_1\leq\left\|\frac{e^{-\beta H}}{\mathcal{Z}_\beta}-\rho\right\|_1. 
\end{align}
From theorem 1 in \cite{perturbstat} it follows that
\begin{equation}
\left\|\frac{e^{-\beta H}}{\mathcal{Z}_\beta}-\rho\right\|_1\leq\kappa(QTQ)\|Q(T-T')Q\|_{1\to1}.  
\end{equation}

As $Q$ is a quantum channel, it follows that $\|Q\|_{1\to 1}\leq1$ and so 
\begin{align}
\|Q(T-T')Q\|_{1\to1}\leq\|T-T'\|_{1\to1}. 
\end{align}
Eq. \eqref{equ:probepsclose} would then follow from $\kappa(QTQ)\leq2+\kappa(T)$.
Note that as $T$ is primitive, we have that $\|T-T_\infty\|<1$, where we use the operator norm.
Also, $QT_\infty Q=T_\infty$ and $Q$ is a projection. Thus
\[
\|QTQ-T_\infty\|\leq \|T-T_\infty\|<1.
\]
As $T$ is an eigenbasis preserving channel, we have that 

\[
Q(T-T_\infty)Q=(T-T_\infty)Q 
\]
and so
\[
(Q(T-T_\infty)Q)^n=Q(T-T_\infty)^nQ. 
\]
We therefore have
\begin{align}\label{equ:neumanninverse}
(\id-(Q(T-T_\infty )Q))^{-1}=\sum\limits_{n=1}^\infty\frac{(Q(T-T_\infty )Q)^n}{n!}=\\
\id-Q+Q\lb\sum\limits_{n=0}^\infty\frac{(T-T_\infty )^n}{n!}\rb Q=\id-Q+Q(\id-(T-T_\infty))^{-1}Q.
\end{align}
As $\|Q\|_{1\to1},\|\id\|_{1\to1}\leq1$ and from \eqref{equ:neumanninverse} we obtain
\begin{align}\label{equ:finalineqchan}
\kappa(QTQ)=\sup\limits_{X\in\M_d,\tr(X)=0}\frac{\|[Q(\id-(T-T_\infty))^{-1}Q+\id-Q](X)\|_1}{\|X\|_1}\leq 
\end{align}
\begin{align*}
\sup\limits_{X\in\M_d,\tr(X)=0}\frac{\|Q(\id-(T-T_\infty))^{-1}Q(X)\|_1+\|(\id-Q)(X)\|_1}{\|X\|_1}\leq2+\kappa(T),
\end{align*}
which completes the proof.
\qed
\hfill\break 

Theorem \ref{thm:stabilitychannel} shows that the algorithm is stable under perturbations of the eigenbasis preserving channel.
The stability for lumpable channels follows by observing that every lumpable channel is in particular eigenbasis preserving.

\subsection{Faulty Phase Estimation}

We will now analyze the errors stemming from faulty phase estimation. 
It is important to differentiate two different types of error
that are caused by the phase estimation procedure. The first type of error comes from the fact that we are only able to obtain an estimate of the energy up to 
$t$ bits from the phase estimation procedure. This leads to round-off errors and may introduce degeneracies.
As discussed in section \ref{sec:lumpingeigenvalues} in theorem \ref{thm:lumpingesigenstates}, algorithm \ref{algo:cftp} is stable against this sort of error.
Moreover, as we will see later, this can even lead to the algorithm being more efficient.

The second kind of error comes from the fact that the phase estimation procedure only gives the correct energy of the state with high probability. 
This can cause
some transitions we record to be corrupted. We will now show that algorithm \ref{algo:cftp}
is stable against this kind of error.

We note that the exact distribution of the outcomes of the phase estimation procedure depends on which version is being used and this is
still a topic of active research~\cite{Svore:2014:FPE:2600508.2600515}.

However, it is reasonable to assume that for any phase estimation routine the distribution of the outcomes will concentrate on the correct output. We will
give bounds in terms of how large this peak is and explicit bounds for the implementation discussed in~\cite[Section 5.2]{nielsen2000quantum}. We now assume we have some rule
to assign the labels based on the outcome of the phase estimation step.
In case we have an $\epsilon$-spectral covering, this might just be a function which assigns the label based on which interval of the covering the outcome of the measurement
belongs to.
Let $X_1\in[d']$ be the random variable which describes which label we assign to the graph after the 
measurement and $Y_1\in\{E_1,\ldots,E_{d'}\}$ the random variable which describes in which eigenspace the system finds itself after the first measurement
at step \ref{step:firstmeasurephase}. 
Let analogously $Y_2$ be the random variable which is distributed according to the probability of each eigenspace at step \ref{step:secondmeasurephase} and
$X_2$ the second label which we assign.
We will now assume that the errors stemming from the phase estimation steps are independent and have the same distribution. That is, given that the system is in a given eigenstate,
the probability distribution of the measurement outcomes is the same in the two steps.
Let the stochastic matrix $\Xi\in\M_{d'}$ be given by
\begin{align*}
\Xi(i,j)=P(Y_1=E_j|X_1=i). 
\end{align*}
Then, given that we have assigned the label $i$ to the graph after step \ref{step:labelgraph} in algorithm \ref{algo:cftp}, the state of the system is described by
\begin{align*}
\rho=\sum\limits_{j=1}^{d'}\Xi(i,j)\frac{P_j}{|S_j|}, 
\end{align*}
where $P_j$ is the projection onto the eigenspace corresponding to $E_j$.
After we apply an eigenbasis preserving channel $T$, the state of the system is described by the state
\begin{align}\label{equ:stateaftermeasurement}
T(\rho)=\sum\limits_{k=1}^{d'}\sum\limits_{j=1}^{d'}\pi(j,k)\Xi(i,j)\frac{P_k}{|S_k|}.
\end{align}
Furthermore, denote by $\Xi'\in\M_{d'}$ the stochastic matrix 
\begin{align*}
\Xi'(i,j)=P(X_2=j|Y_2=E_i). 
\end{align*}
From Eq. \eqref{equ:stateaftermeasurement} it then follows that the probability that the second label is $l$ given that the first label
was $i$ is
\begin{align}\label{equ:transitionmatrix}
P(X_2=l|X_1=i)=\sum\limits_{k,j,l=1}^{d'}\Xi'(k,l)\pi(j,k)\Xi(i,j). 
\end{align}
From Eq. \eqref{equ:transitionmatrix} it is clear that the transition matrix for the labels is given by
\begin{align}\label{equ:righttransition}
 \pi'=\Xi'\pi\Xi
\end{align}
when we have faulty phase estimation.

As mentioned before, we expect $P(X_1=i|Y_1=E_j)\simeq\delta_{i,j}$, that is, that the distribution peaks around the right outcome.
To quantify this we define
\begin{align}\label{equ:definxi}
 \xi=\min\limits_{i\in[d']}\Xi_{i,i}
\end{align}
and $\xi'$ analogously. We then have

\hfill\break 
\begin{lem}
Let $\xi$ and $\xi'$ be defined as above and $\{F_i\}_{i\in I}$ a POVM.  For a primitive lumpable channel $T:\M_d\to\M_d$
for a Hamiltonian $H$ and inverse temperature $\beta>0$, let $p(i)=\tr\lb F_i\frac{e^{-\beta H}}{\mathcal{Z}_\beta}\rb$ and $p'(i)$ be the probability of observing $F_i$ at the output of 
algorithm \ref{algo:cftp} with faulty phase estimation.
Then	
\begin{align*}
\|p-p'\|_1\leq 1-\xi'+2\lb\kappa(T)+2\rb((1-\xi\xi'+(1-\xi)\xi'+\xi(1-\xi')+(1-\xi)(1-\xi')). 
\end{align*}

\end{lem}

\hfill\break 
\noindent\ignorespacesafterend
\textbf{Proof:} As discussed in Eq. \eqref{equ:righttransition}, the transition matrix for the observed energy labels is given by $\pi'=\Xi'\pi\Xi$.
It easily follows from the definition of $\xi$ and $\xi'$ that 
\begin{align*}
\Xi=\xi\one+(1-\xi)\tilde{\Xi},\\
\Xi'=\xi'\one+(1-\xi')\tilde{\Xi}',
\end{align*}
where $\tilde{\Xi}$ and $\tilde{\Xi}'$ are again stochastic matrices. We may therefore write
\begin{align*}
 \pi'=\xi\xi'\pi+\xi(1-\xi')\tilde{\Xi}'\pi+(1-\xi)\xi'\pi\tilde{\Xi}+(1-\xi)(1-\xi')\tilde{\Xi}'\pi\tilde{\Xi}.
\end{align*}
This transition matrix will still be primitive for $\xi$ and $\xi'$ sufficiently large. 
Let $\mu$ be the stationary distribution of $\pi$ and $\mu'$ the one of $\pi'$. 
Observe that, as $\kappa(\pi)\leq\kappa(QTQ)$, we may use the bound $\kappa(QTQ)\leq2+\kappa(T)$ from the proof of theorem \ref{thm:stabilitychannel} and obtain

\begin{align*}
&\|\mu-\mu'\|_1\leq \\
\lb\kappa(T)+2\rb \|\xi\xi'\pi+\xi(1-\xi')\tilde{\Xi}'\pi+&(1-\xi)\xi'\pi\tilde{\Xi}+(1-\xi)(1-\xi')\tilde{\Xi}'\pi\tilde{\Xi}-\pi\|_{1\to1} \\ 
\leq2\lb\kappa(T)+2\rb((1-\xi\xi'+&(1-\xi)\xi'+\xi(1-\xi')+(1-\xi)(1-\xi')).
\end{align*}
Here we have used that $\|\pi\|_{1\to1}\leq1$ for a stochastic matrix $\pi$. 
At the output of the algorithm, we would be measuring the POVM on the state $\rho'=\sum_{i=1}^{d'}\mu'(i)\frac{P_i}{|S_i|}$ if no error occurs at step \ref{step:firstmeasurephase} of algorithm 
\ref{algo:cftp}.
But as an error might occur when we try to identify a given eigenstate, we will be measuring the state $\rho_{EM}=\sum_{i=1}^{d'}(\Xi' \mu')(i)\frac{P_i}{|S_i|}$.
By the definition of $\xi'$, we have $\Xi'\mu'=\xi'\mu'+(1-\mu')\tilde{\Xi'}\mu'$. We will measure the POVM on the state
\begin{align}
\rho_{EM}=\xi'\rho'+(1-\xi')\rho''. 
\end{align}
Here $\rho''$ is some density matrix.
It then follows that
\begin{align*}
\left\|\rho_{EM}-\gibbs\right\|_{1}\leq1-\xi'+2\lb\kappa(T)+2\rb((1-\xi\xi'+(1-\xi)\xi'+\xi(1-\xi')+(1-\xi)(1-\xi')). 
\end{align*}
The claim then follows from the variational expression for the trace distance as in the proof of theorem \ref{thm:stabilitychannel}.
\qed
\hfill\break 

Using Bayes' rule it is possible to express the entries of the matrix $\Xi$ in terms of those of $\Xi'$, which are more readily accessible. We have
\begin{align}\label{equ:bayesoutcome}
\Xi(i,j)=P(Y_1=E_j)\frac{P(X_1=i|Y_1=E_j)}{\sum\limits_{l=1}^{d'}P(X_1=i|Y_1=E_{l})P(Y_1=E_{l})}. 
\end{align}
As the initial state is the maximally
mixed one, we have that $P(Y_1=E_j)=|S_j|d^{-1}$.
From this discussion it follows that:
\hfill\break 
\begin{thm}\label{thm:estimatexi}
Let $\xi$ be defined as in Eq. \eqref{equ:definxi}. Then 
\begin{align*}
\xi=\min\limits_{j\in[d']}\frac{P(X_1=j|Y_1=E_j)}{|S_j|^{-1}\sum\limits_{l=1}^{d'}P(X_1=j|Y_1=E_{l})|S_l|}.
\end{align*}
\end{thm}
\hfill\break 
\noindent\ignorespacesafterend
\textbf{Proof:} See the discussion above.
\qed
\hfill\break 

This shows that the algorithm is stable if we do not have eigenvalues that we can misidentify with considerable probability
and s.t. the degeneracy levels are of different order.

We now give estimates of $\xi$ and $\xi'$ for the implementation of phase estimation considered in~\cite[Section 5.2]{nielsen2000quantum} 
in case
we have an $\epsilon$-spectral covering of the Hamiltonian or know that different eigenvalues are $\epsilon$ apart.
In~\cite{nielsen2000quantum} it is shown that if we use
\begin{align*}
t\geq n+\log\lb2+(2\delta)^{-1}\rb 
\end{align*}
qubits to perform phase estimation, then we obtain $E_i$ accurate to $n$ bits with probability at least $1-\delta$. 
This implies that $\xi'\geq1-\delta$.
To estimate $\xi$ we need to control the terms of the form $P(X_1=i|Y_1=E_j)$ for $i\not=j$. 
To this end we define 
\begin{align}\label{equ:defiDelta}
\Delta(i,j)=\inf\{|2^tx-2^ty \mod 2^t||(x,y)\in A(i,j)\}
\end{align}
with
\begin{align*}
A(i,j)=\{x\in\sigma(H)\cap (E_i-\epsilon,E_i+\epsilon)\}\times\{y\in(E_j-\epsilon,E_j+\epsilon)\}.   
\end{align*}
\begin{lem}\label{lem:prob}
Let $H\in\M_d$ be a Hamiltonian and $\{e_1,\ldots,e_{d'}\}$ be an $\epsilon$-spectral covering of it. 
Suppose we implement phase estimation for $H$ using $t$ qubits. Then
\begin{align*}
P(X_1=j|Y_1=e_i)\leq\frac{2^{t+1}\epsilon+1}{\Delta(i,j)^2}
\end{align*}
for $j\not=i$ and $\Delta_{i,j}$ defined as in Eq. \eqref{equ:defiDelta}.
\end{lem}
\hfill\break 
\noindent\ignorespacesafterend
\textbf{Proof:} In ~\cite[Section 5.2]{nielsen2000quantum}  it is shown that given that the eigenstate of the system is $E_i$, we have that the probability that
the observed outcome is $E$ is bounded by 
\begin{align*}
|2^t(E_i-E) \mod 2^t|^{-2}.  
\end{align*}

For any point of the spectrum of $H$ in $(e_i-\epsilon,e_i+\epsilon)$ and for a point in
$E\in(e_j-\epsilon,e_j+\epsilon)$ we have that 
$|2^t(e_i-E) \mod 2^t|^{-2}\geq\Delta(i,j)$. There are at most $2\epsilon2^t+1$ possible outcomes that lie in the interval $(e_j-\epsilon,e_j+\epsilon)$.
We therefore have
\begin{align*}
P(X_1=j|Y_1=e_i)\leq\frac{2^{t+1}\epsilon+1}{\Delta(i,j)^2}.\qed
\end{align*} 
\hfill
Note that we have $2^t\leq\Delta(i,j)$, so the probability of misidentifying the labels goes to zero exponentially fast with the number of qubits for fixed $\epsilon$.
We then obtain for $\xi$ and $\xi'$:
\hfill\break 
\begin{cor}\label{cor:finalestimatexi}
Let $H\in\M_d$ be a Hamiltonian and $\{e_1,\ldots,e_{d'}\}$ be an $\epsilon$-spectral covering of it with $\epsilon\geq 2^{-n}$. 
Suppose we implement phase estimation for $H$ using $t\geq n+1+\log\lb2+(2\delta)^{-1}\rb$ qubits. 
Then $\xi'\geq1-\delta$ and
\begin{align}\label{equ:estimatexifinal}
\xi\geq \min\limits_{j\in[d']}\frac{1-\delta}{1-\delta+(2^{t+1}\epsilon+1)|S_j|^{-1}\sum\limits_{l\not=j}\Delta(j,l)^{-2}|S_l|}. 
\end{align}
\end{cor}
\hfill\break 
\noindent\ignorespacesafterend
\textbf{Proof:} As $\epsilon\geq 2^{-n}$ and with probability at least $1-\delta$ we will obtain an output which is accurate up to $n+1$ bits, with probability
at least $1-\delta$ we will correctly identify in which element of the covering we are from the output. From this, it follows that $\xi'\geq1-\delta$.
As the function $(x,y)\mapsto\frac{x}{x+y}$ is monotone increasing in $x$ and decreasing in $y$ for $x,y>0$, we obtain Eq. \eqref{equ:estimatexifinal}
by inserting the bound on $\xi'$ and the result of Lemma \ref{lem:prob} into the expression we derived for $\xi$ in theorem \ref{thm:estimatexi}.
\qed
\hfill\break 

Corollary \ref{cor:finalestimatexi} clarifies that the algorithm requires a larger number of qubits to be reasonably
stable if we have close $E_j$ and $E_{j'}$ s.t. $|S_j|\ll |S_{j'}|$. The converse is also true; if $|S_j|\backsimeq |S_{j'}|$ for all $j,j'$ the algorithm is
already stable with a small precision.

\section{Expected run-time, Memory Requirements and Circuit Depth}\label{sec:efficiency}

We will now address the expected run-time of algorithm \ref{algo:cftp}. To this end, we will only consider the number of calls
of the phase estimation and eigenbasis preserving or lumpable channel and not the necessary classical post-processing, as we consider the quantum routines
the more expensive resources.

In \cite[Theorem 5]{Propp1998170}, it was shown that the expected time to obtain a sample using algorithm \ref{algo:cftppropp} is $\mathcal{O}(t_{\textrm{mix}}|S|^2)$ steps, where again $t_{\textrm{mix}}$ is the time such that the chain 
is $e^{-1}$ close to stationarity and $|S|$ the size of our state space. In the case of Hamiltonians with degenerate spectrum $t_{\textrm{mix}}$ will denote the mixing time of the classical lumped chain induced by the lumpable channel (see Eq. 
\eqref{equ:transitionlumpable}).

Recall that $d'$ denotes the number of distinct eigenvalues of the Hamiltonian or the size of the $\epsilon$-spectral covering being used.
That is, $d'$ is just the number of different labels of the graph.
We will say that a column indexed by $k\in-\N$ of $G$ is complete if $\forall i\in[d']$ $G(k,i)\not=0$. 
In \cite{Propp1998170}, it is shown that we need to complete on average $\mathcal{O}(t_{\textrm{mix}}d')$ columns of the
graph $G$ before the labels on a column become constant. 
As each step to complete a column needs $\mathcal{O}(d')$ calls of $\mathbf{RandomSuccessor}$, this leads to a total of $\mathcal{O}(t_{\textrm{mix}}{d'}^2)$ 
calls of $\mathbf{RandomSuccessor}$.
The dynamics in the eigenbasis of $H$ is classical, so we may use the exact same reasoning to conclude that we will need an expected 
$\mathcal{O}(t_{\textrm{mix}}d')$ 
number of complete columns until we obtain one perfect sample.

But in our case, we may need more uses of the channel and phase estimation, as we may not prepare an arbitrary eigenstate of $H\in\M_d$ which might be necessary 
to complete a column deterministically. We will denote the expected number of measurements necessary to complete a column by $\phi(H)$
and in theorem \ref{thm:exactnumberofsamples} in the Appendix \ref{app:numbermeasur} we give an explicit expression for this quantity.

In Appendix \ref{app:numbermeasur} we prove bounds on $\phi(H)$ for various cases of interest and remark that in the worst case, namely Hamiltonians with a non-degenerate spectrum, $\phi(H)=\bigO{d\log(d)}$.
Preparing the initial states probabilistically does not significantly
change the overall efficiency of the algorithm, as illustrated by the next theorem.
\hfill\break 
\begin{thm}\label{thm:runtime}
Let $T:\M_d\to\M_d$ be a lumpable quantum channel for a Hamiltonian $H$ at  inverse temperature $\beta>0$ with mixing time $t_{\textrm{mix}}$.
Then the expected number of steps until algorithm \ref{algo:cftp} returns a perfect sample is $\mathcal{O}(t_{\textrm{mix}}d'\phi(H))$.
\end{thm}
\hfill\break 
\noindent\ignorespacesafterend
\textbf{Proof:} We will need an average of $\phi(H)$ measurements to complete a column.
From the result \cite[Theorem 5]{Propp1998170} we know that we will need an expected number of $\mathcal{O}(t_{\textrm{mix}}d')$ number of complete columns to obtain a sample.
As the number of measurements needed to complete a column and complete columns to obtain a sample are independent, we have an expected $\mathcal{O}(t_{\textrm{mix}}d'\phi(H))$ number of steps
to obtain a sample.\qed
\hfill\break

It should be clear from theorem \ref{thm:runtime} that algorithm \ref{algo:cftp} is considerably less efficient than other algorithms such as quantum Metropolis \cite{temme2011quantum} if we are willing to settle for an approximate sample
for Hamiltonians with a non-degenerate spectrum. In this case we have $d'=d$ and $\phi(H)=\bigO{d\log(d)}$, giving a total complexity of $\mathcal{O}(t_{\textrm{mix}}d^2\log(d))$ in the worst case.
After all, to obtain a sample that is $e^{-1}$ close in trace distance to the Gibbs state, one only needs $t_{\textrm{mix}}$ steps of the Metropolis algorithm instead of the $\mathcal{O}(t_{\textrm{mix}}d^2\log(d))$ needed for CFTP.
Therefore, it is important to stress again that these algorithms are very different in nature. Algorithm \ref{algo:cftp} provides us with perfect, not approximate samples, and it is the first algorithm of this form
for quantum Gibbs states to the best of our knowledge. It provides a certificate that we are indeed sampling from the right distribution when it terminates, while most other algorithms require some mixing time bounds to obtain a 
sample that can be considered close to the target distribution. Moreover, it only requires us to be able to implement one step
of the chain.

However, for the case of Hamiltonians with a highly degenerate spectrum, our algorithm is efficient, as is illustrated by the next theorem:
\hfill\break 
\begin{thm}\label{thm:runtime2}
Let $T:\M_d\to\M_d$ be a lumpable quantum channel for a Hamiltonian $H\in\M_d$ at  inverse temperature $\beta>0$ with mixing time $t_{\textrm{mix}}$.
Moreover, assume that we have $d\leq|S_i| r(d)$ for some function $r:\R\to\R$ and all eigenspaces $S_i$. 
Then the expected number of steps until algorithm \ref{algo:cftp} returns a perfect sample is $\mathcal{O}(t_{\textrm{mix}}r(d)^2\log(r(d)))$.
\end{thm}
\hfill\break 
\noindent\ignorespacesafterend
\textbf{Proof:} From $d\leq|S_i| r(d)$ it follows that $d'\leq r(d)$.
It follows from theorem \ref{thm:bounduniform} that we will need an average of $\bigO{r(d)\log(r(d))}$ measurements to complete a column.
From the result \cite[Theorem 5]{Propp1998170} we know that we will need an expected number of $\mathcal{O}(t_{\textrm{mix}}d')$ number of complete columns to obtain a sample.
As the number of measurements needed to complete a column and complete columns to obtain a sample are independent, we have an expected $\mathcal{O}(t_{\textrm{mix}}r(d)^2\log(r(d)))$ number of steps
to obtain a sample.
\qed
\hfill\break 

In particular, for the cases $r(d)=c$ for some $c\in\R$, which corresponds to having eigenspaces with a degeneracy proportional to the dimension, we have that we only need $\bigO{t_{\textrm{mix}}}$
steps to obtain a perfect sample. That is, the time necessary to obtain perfect samples with our algorithm and approximate ones are the same up to a constant factor. 
Slightly more generally, for $r(d)=c\log(d)^m$ our algorithm still has a polylogarithmic runtime and is efficient. 
Admittedly such level of degeneracy is not usual for Hamiltonians of physical relevance. One could use the strategy discussed in section \ref{sec:lumpingeigenvalues} 
and still obtain certifiably good samples by lumping together eigenvalues that are close. Moreover, as we will only need to run a $r(d)$ dimensional version of classical CFTP, 
the classical part of the algorithm will be efficient.

Although the worst case $\mathcal{O}(t_{\textrm{mix}}d^2\log(d))$ scaling is prohibitive for applications, this is still more efficient than explicitly diagonalizing $H$ as long as $t_{\textrm{mix}}\log(d)=\mathcal{O}(d^{\omega-2})$.
Here $2<\omega<2.373$ is the optimal exponent of matrix multiplication, which has the same complexity as diagonalization~\cite{demmel2007fast}. That is, as long as approximate sampling is efficient,
obtaining perfect samples is faster than diagonalizing even in the worst case.

We now analyze the circuit depth and memory requirements to obtain a sample.
\hfill\break 
\begin{thm}
Let $C_{PT}$ and $C_{T}$ be the circuit depth needed to implement the phase estimation for $H$ and the eigenbasis preserving channel, respectively. 
Then one needs to implement a quantum circuit of depth $\mathcal{O}(C_{PT}+C_{T})$ to obtain a sample and moreover an expected $\mathcal{O}(\phi(H))$ classical memory.
\end{thm}
\hfill\break 
\noindent\ignorespacesafterend
\textbf{Proof:} The circuit length part follows easily from just going through the steps of algorithm \ref{algo:cftp}, as to label the new vertex we need to implement two phase estimation steps and apply the eigenbasis preserving channel once. 

To see the that we only need $\mathcal{O}(\phi(H))$ classical memory, notice that 
we only need to store the information contained in the last complete column to perform the later steps. This is because it contains all possible labels for future columns. By corollary \ref{cor:numbermeasur}, we
have that the expected number of labels we obtain before completing a column is $\mathcal{O}(\phi(H))$, and so we need a total classical memory of size $\mathcal{O}(\phi(H))$.
\qed
\hfill\break

The quantum part of algorithm \ref{algo:cftp} can be easily parallelized, as we could use different quantum computers feeding a classical computer with valid transitions.
Note that the classical resources necessary to run the algorithm are also not very large in the cases in which we have a highly degenerate spectrum, as discussed before.

\section{Adapting other Variations of CFTP}
In \cite{Propp1998170} the authors discuss other variations of CFTP that can be more efficient, such as the cover time CFTP algorithm.
We will not discuss in detail how to adapt these other proposals, but it should be straightforward to do so from the results in the last sections.
In this section, we will just mention the main ideas.
Note that the only thing necessary to implement all these variations is a valid $\mathbf{RandomSuccessor}$ function and 
the outputs of the measurements in steps \ref{step:firstmeasurephase} and \ref{step:secondmeasurephase} of algorithm \ref{algo:cftp} do exactly that.
This information could then be fed to a classical computer running a variation of CFTP.
The only difference to the classical case is that we may not choose arbitrary initial states, but do so probabilistically.
However, by waiting until each initial state is observed, we may circumvent this and do not have a significant overhead by the result of corollary
\ref{cor:numbermeasur}.

For some variations of CFTP, like again the cover time CFTP, one needs to iterate $\mathbf{RandomSuccessor}$. This is also straightforward.
If we want to obtain a given number of iterations of $\mathbf{RandomSuccessor}$, we just apply an eigenbasis preserving or lumpable channel $T$ to the first register, repeated 
by a phase estimation step and a measurement in the computational basis. We then repeat this procedure to obtain the iterations.

One could then repeat the analysis done in this section and see that the run-time is again of the same order of magnitude as the classical version of the CFTP algorithm and obtain a 
perfect sampling algorithm with a run-time proportional to the cover time of the lumped chain.

\section{Conclusion and Open Problems}
We have shown how to adapt perfect sampling algorithms for classical Markov chains to obtain perfect samples of
quantum Gibbs states on a quantum computer. These algorithms have an average run-time which in the worst case is similar to their classical counterparts.
For highly degenerate Hamiltonians this algorithm gives an efficient sampling scheme and in the extreme case of having degeneracies proportional to the dimension,
the time required to sample perfectly is even proportional to the time necessary to obtain an approximate sample. In these cases, the classical post-processing required
can be done efficiently. We showed how to increase the efficiency of the sampling scheme and still obtain certifiably good samples by lumping close eigenstates together.
We argue that one of its main advantages is its short circuit depth.
We show that the algorithm is stable under noise in different steps of the implementation.
It would be interesting to find sampling applications or models that satisfy the conditions under which our algorithms are efficient.
Moreover, it would be worthwhile to investigate if there is a class of models to which we can tailor the perfect sampling algorithms to be efficient, as was done with success
for attractive spin systems~\cite{propp1996exact}.

\section*{Acknowledgments}
D.S.F. would like to thank Michael M. Wolf for raising the question of how to perfectly sample from a quantum Gibbs state and the constant support
and Andreas Bluhm for helpful discussions.
D.S.F acknowledges support from the graduate program TopMath of the Elite Network of Bavaria, the 
TopMath Graduate Center of TUM Graduate School at Technische Universit\"{a}t M\"{u}nchen, the Deutscher Akademischer 
Austauschdienst(DAAD) and by the Technische Universit\"at M\"unchen – Institute for Advanced Study,
funded by the German Excellence Initiative and the European Union Seventh Framework
Programme under grant agreement no. 291763.

\nonumsection{References}
\vspace{-1em}
\noindent
\bibliographystyle{unsrt}
\bibliography{bibliography}

\begin{appendices}

\section{Expected number of observations to sample all possible outputs}\label{app:numbermeasur}
To estimate the expected run-time of algorithm \ref{algo:cftp}, we need to determine how often, on average, we must measure projections
 $\{P_i\}_{1\leq i \leq d'}$ on the state $\frac{\one}{d}$
until we observe all possible outcomes $i$. Here the $P_i$ correspond to projections onto different eigenspaces of the Hamiltonian Gibbs state $\frac{e^{-\beta H}}{\mathcal{Z}_\beta}$
we are trying to sample from.
The number of measurements corresponds to the time necessary to complete a column in algorithm \ref{algo:cftp} and will be denoted by $\phi(H)$.

One can see that this corresponds to the classical problem of determining how many coupons one must collect to obtain at least one of each, the coupon collector problem~\cite{neal_2008}.
But in our case, we have unequal probabilities for the coupons or outcomes, as the probability $q(i)$ of observing $i$ is given by 
\begin{align}\label{equ:probobsi}
q(i)=\tr\lb P_i\frac{\one}{d}\rb=\frac{|S_i|}{d}. 
\end{align}
\hfill\break 
\begin{thm}[Coupon collector with unequal probabilities]\label{thm:exactnumberofsamples}
Let $q\in\R^{d'}$ be the probability distribution of the measurement outcomes as in Eq. \eqref{equ:probobsi}. Let $Y$ be the random variable given by the number of measurements
necessary to observe all possible outputs.
Then
\begin{align}\label{equ:expectedtimecoupon}
\phi(H)=E(Y)=\sum\limits_{j=0}^{d'-1}(-1)^{d'-1-j}\sum\limits_{|J|=j}(1-Q_J)^{-1}, 
\end{align}
where $Q_J$ is defined as $Q_J=\sum_{i\in J}q(i)$ for $J\subset[d']$ and the second sum is over all subsets of size $j$.
\end{thm}
\noindent\ignorespacesafterend

\hfill\break 
\noindent\ignorespacesafterend
\textbf{Proof:}  We refer to~\cite[Corollary 4.2]{flajolet1992birthday} for a proof.
\qed
\hfill\break 

Although the expression in Eq. \eqref{equ:expectedtimecoupon} is exact, its asymptotic scaling is not clear from it.
Therefore, we show the following bound which is more directly accessible.
\hfill\break 
\begin{thm}\label{thm:bounduniform}
Let $q\in\R^{d'}$ be the probability distribution of the measurement outcomes as in Eq. \eqref{equ:probobsi}. Let $Y$ be the random variable given by the number of measurements
necessary to observe all possible outputs. Moreover, assume that $d\leq|S_i| r(d)$ for some $r(d):\R\to\R$. Then
\begin{align*}
\phi(H)=E(Y)\leq r(d)\varphi_{d'}=\mathcal{O}\lb r(d)\log\lb r(d)\rb\rb, 
\end{align*}
where $\varphi_{d'}=\sum\limits_{l=1}^{d'}\frac{1}{l}$.
\end{thm}
\hfill\break 
\noindent\ignorespacesafterend
\textbf{Proof:} We mimic the proof of the classical result for the coupon collector problem with uniform probability distribution. 
Denote by $t_l$ the expected time to collect a new coupon after $l-1$ have been collected.
We then have 
\begin{align*}
E(Y)=\sum\limits_{l=1}^{d'}E(t_l). 
\end{align*}
We clearly have $E(t_1)=1$. 
Define $k(d)=\frac{d}{r(d)}$.
For $l\geq2$, note that as we have that we get each coupon with probability at least $\frac{k(d)}{d}$, the probability of getting a new coupon after having collected
$l-1$ is at least $(d'-l+1)\frac{k(d)}{d}$. From this, it follows that $E(t_l)\leq\frac{d}{k(d)}\lb d'-l+1\rb^{-1}$ and so
\begin{align*}
E(Y)\leq1+\frac{d}{k(d)}\sum\limits_{l=2}^{d'}\frac{1}{d'-l+1}\leq\frac{d}{k(d)}\sum\limits_{l=1}^{d'}\frac{1}{l}.  
\end{align*}
The claim follows from observing that $\varphi_d=\mathcal{O}(\log(d))$ and that $d'\leq r(d)$.
\qed
\hfill\break 

From this, it is easier to get estimates for cases that might be of interest. Here we collect the bounds for the extreme cases of highly degenerate spectra, that is, with each eigenspace having
dimension $\Omega\lb d\log(d)^{-m}\rb$ for $m\in\N$ and the non-degenerate case.
\hfill\break 
\begin{cor}\label{thm:columnverydegenerate}
Let $q\in\R^{d'}$ be the probability distribution of the measurement outcomes as in Eq. \eqref{equ:probobsi}. Let $Y$ be the random variable given by the number of measurements
necessary to observe all possible outputs. Moreover, assume that $|S_i|\geq c\frac{d}{\log(d)^m}$ for some $c\in\R$ and all $i\in[d']$. Then
\begin{align*}
\phi(H)=E(Y)\leq \frac{\log(d)^m}{c}\varphi_{c^{-1}\log(d)^m}=\mathcal{O}\left(\log\lb d\rb^m\log\left[ m\log(d)\right]\right).
\end{align*}
\end{cor}
\hfill\break 
\noindent\ignorespacesafterend
\textbf{Proof:} Just take $r(d)=c^{-1}\log(d)^m$ in theorem \ref{thm:bounduniform}.
\qed
\hfill\break

\hfill\break 
\begin{cor}\label{cor:numbermeasur}
Let $q\in\R^{d'}$ be the probability distribution of the measurement outcomes as in Eq. \eqref{equ:probobsi}. Let $Y$ be the random variable given by the number of measurements
necessary to observe all possible outputs. Moreover, assume that $q(i)=\frac{1}{d}$. Then
\begin{align*}
\phi(H)=E(Y)=\mathcal{O}\lb d\log\lb d\rb\rb
\end{align*}

\end{cor}
\hfill\break 
\noindent\ignorespacesafterend
\textbf{Proof:} Just take $r(d)=d$ in theorem \ref{thm:bounduniform}.
\qed
\hfill\break 

That is, we might go from a constant number of samples necessary to complete a column in the case of degeneracies proportional to the dimension to a scaling like $d\log(d)$.
One should note that applying the bound in corollary \ref{cor:numbermeasur} to analyze the runtime of algorithm \ref{algo:cftp} probably leads to bounds that are too pessimistic for spectra that are not very degenerate. 
To see why this is the case, note that in algorithm \ref{algo:cftp} we do not discard measurements outcome we have already observed, but rather use them to complete other columns, which we do not take into account in this analysis.

\end{appendices}

\end{document}